# Reliable Communication in Wireless Body Area Sensor Network for Health Monitoring

Saeid Bahanfar[1], Ladan Darougaran[2], Helia Kousha[3] and Shahram Babaie[4]

[1] Department of Computer Engineering, Tabriz Branch, Islamic Azad University, Tabriz, Iran

[2] Department of Computer Engineering, Tabriz Branch, Islamic Azad University, Tabriz, Iran

[3] Department of Computer Engineering, Tabriz Branch, Islamic Azad University, Tabriz, Iran

[4] Department of Computer Engineering, Tabriz Branch, Islamic Azad University, Tabriz, Iran

**Abstract**

Now days , interests in the application of Wireless Body Area Network (WBAN) have grown considerably. A number of tiny wireless sensors, strategically placed on the human body, create a wireless body area network that can monitor various vital signs, providing real-time feedback to the user and medical personnel. This communication needs to be energy efficient and highly reliable while keeping delays low. In this paper we present hardware and software architecture for BAN and also we offer reliable communication and data aggregation.

## 1. Introduction

Nowadays, one of the major applications of wireless sensor networks is environmental monitoring. In these networks, an abundance of sensors is scattered around to collect and retrieve environmental data. A new use of sensor networks can be found in the area of wearable health monitoring. Carefully placing sensors on the human body and wirelessly connecting them to monitor physiological parameters like heartbeat, body temperature, motion et cetera is a promising evolution. This system can reduce the enormous costs of patients in hospitals as monitoring can occur real-time, over a longer period and at home [1, 2]. This type of network is called a Wireless Body Area Network (WBAN) or Wireless Body Sensor Network (WBSN) [3,5,7]. A WBAN consists of several sensors and possibly actuators equipped with a radio interface. Each WBAN has a sink or personal server such as a PDA[4], that receives all information from the sensors and provides an interface towards other networks or medical staff. Connecting health monitoring sensors wirelessly improves comfort for patients but induces a number of technical challenges like coping with mobility and the need for increased reliability. An important requirement in WBANs is the energy efficiency of the system. The sensors placed on the body only have limited battery capacity or can scavenge only a limited amount of energy from their environment [6,8]. In this paper we offer Reliable communication in wireless body area sensor network for health monitoring. We organized rest of the paper section2 about architecture and section3 we offered a way for error detection, section4 explanation about communication and the end section5 is conclusion.

## 2. Architecture

Was designed so that the sensor nodes that are small and use the batteries so that their lives for a long time. Nodes crude weak signals from the human body are collected. The most common physiological signals, (Sp) pulse rate, respiration rate, spirometry, ECG, body temperature, blood gas levels, cardiac output, blood pressure.
Methods BAN this case acts in the human body sensors (with HUB local) player is the least disturbance for the individual to assume sensors along the blood flow in the human body vital signs his measurements are always these symptoms their control and in case of any problem in the system of human body in different ways that information to emergency centers after appropriate physician information necessary measures to person and gives the desired information as possible from the problems caused by this disorder avoids.

2.1 Architecture (for cluster head sensor)

Ordering the front view of the sensor inside the body if the body is one area such as a cluster area and imagine for a CH Cluster let out a task with environmental data collection and programs to run interrupt. Figure1 explains ordering of sensors in body.

2.2 **Architecture offer**

The architecture under a hardware interrupt occurs (figure 2). When an interrupt occurs (low level) under the program runs

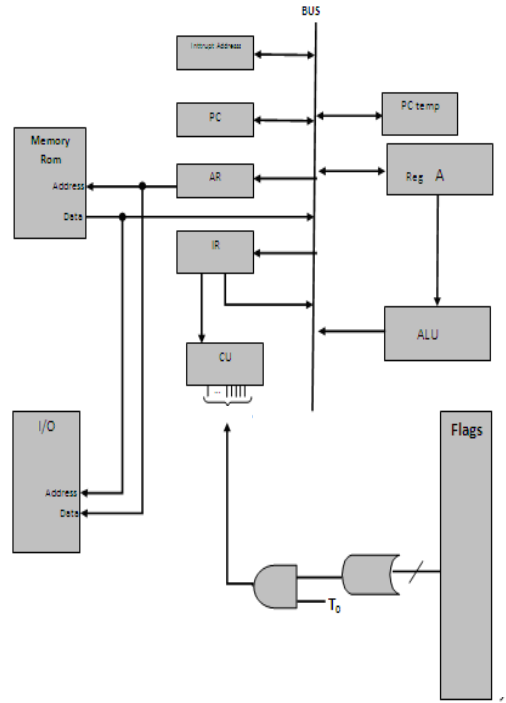

Fig. 2. Architecture

When an interrupt occurs under the program interrupt (high level) is performed (described later in different scenarios will be.) First, CH, data is sent with the normal template and immediately (after the default $100_{ms}$ ) will go to your inbox if you did not receive a message to the following steps. First, the second data format to be sent to other cell phone I do not give two seconds if you did not answer (ACK) data format and frequency of the third device(phone) sends to the default 20 seconds to wait and interrupt is removed ( RetI) Now if you interrupt the program run out again in a state of emergency occurs interrupt instruction interrupt occurs again running this operation to be performed when a state of emergency is removed or power to all sensors (sensor up their efforts person to survive will do).

$time_0 = 24_h$
$time_{01} = 1_h$
$time_1 = 100_{ms}$
$time_2 = 2_s$
$time_3 = 20_s$

A state of emergency occurs interrupt occurs
**Interrupt:**

Send (format data1)
Standby ( $time_1$ )
If (check inbox == false)

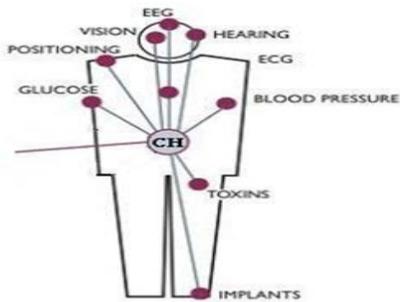

Fig. 1. Body Area Network

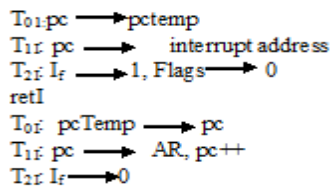

$T_{01}: pc \longrightarrow pctemp$
$T_{1f}: pc \longrightarrow$ interrupt address
$T_{2f}: I_f \longrightarrow 1, Flags \longrightarrow 0$
retI
$T_{0f}: pcTemp \longrightarrow pc$
$T_{1f}: pc \longrightarrow AR, pc++$
$T_{2f}: I_f \longrightarrow 0$

(
    Send (format data2)
    Standby ($time_2$)
)
Else
(If (check inbox == true)
    (Empty (inbox)
        Standby ($time_3$)
        Return interrupt
    )
    Else
     (Send (format data3)
Standby ($time_3$)
Return interrupt
    )
)

## 3. Error detection

BAN ideas in this article that acts this way in all parts of the human body has been used sensors.
Here is a problem that may occur is a condition that the sensor is defective. Data from the body feels is not valid.

### 3.1. Error detection:

in each sensor, we built a neural network. Each sensor data in addition to those who feel their body has data of its neighboring sensors will also receive Function neural network training with the sensor data received and produced a data is that the sense data with data by the sensor is relevant to compare these two data error rate is calculated with the area defined as the maximum possible error is calculated, comparison is done if the error rate obtained is less than the amount defined by the sense data is correctly declared and will be sent to CH. For the sensor itself with the changing environment to adapt its education and only stage production is not the sensor, the neural network re-sense data correctly with the existing training and this cause we update training neural networks and sensors will lead to increase the reliability of neural network will be built. But otherwise, if the error rate range is defined more by a sense of false data is detected The first approach that, the actual range for each sensor in the human body that may happen and if we define sensor is faulty sense of data that is outside this range is timer: The timer mode, the sensor will begin to countdown if this period (until the overflow timer has not happened) data generated with data by neural network prediction is contrary it can be concluded that the sensor is disabled (After this period, individual data base based on the desired message that one of your sensor sends damaged). If before the timer overflow occurs sense data with predicted data is equal to the primary mode timer returns and is diagnosed, the sensor is not defective and the immediate effect of these uncertainties there is a contradiction.

Each sensor as Figure (3) shows includes the following sections.

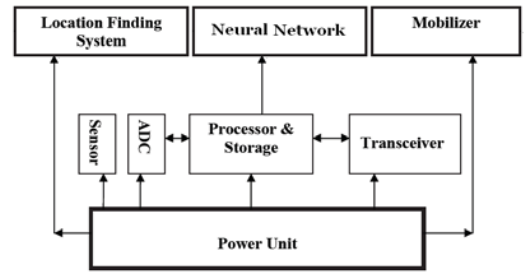

Fig. 3 Interior Sensor.

Performance of neural networks is that sensors within his duty is to determine the status of sensors (for example, send a zero element means conflict between sense data and data predicted by neural network is a means to send the same data is two. Posts by 1111101111 CH in the Status field sensors means that there are 10 sensors that sense data by the fifth sensor data generated neural network with the same sensor is different. Position Sensor in place bit location in the body determines the sensor can be that their data is important to have a bit space is valuable in this case, if the sensor is more important from the database can be displayed with high priority. the range data that each sensor can send the number of bits that send data for each sensor will be compared will be different), and vital signs that are sent with authentication.

When the cluster head receives data using a threshold to detect conditions normal or abnormal. In the normal case production of all sensor data in the threshold is an unusual case but the two conditions occur eg If less than three sensors in the threshold of any CH situation semi-critical diagnosis and a message to all broadcast to the other sensors to based on that they data to CH to send and CH after receiving their data to send the database. In critical case condition for more than three sensors placed in the threshold in this case for CH when all the broadcast message does not die in a state of crisis because the sensor data directly to other emergency and need not be informed.

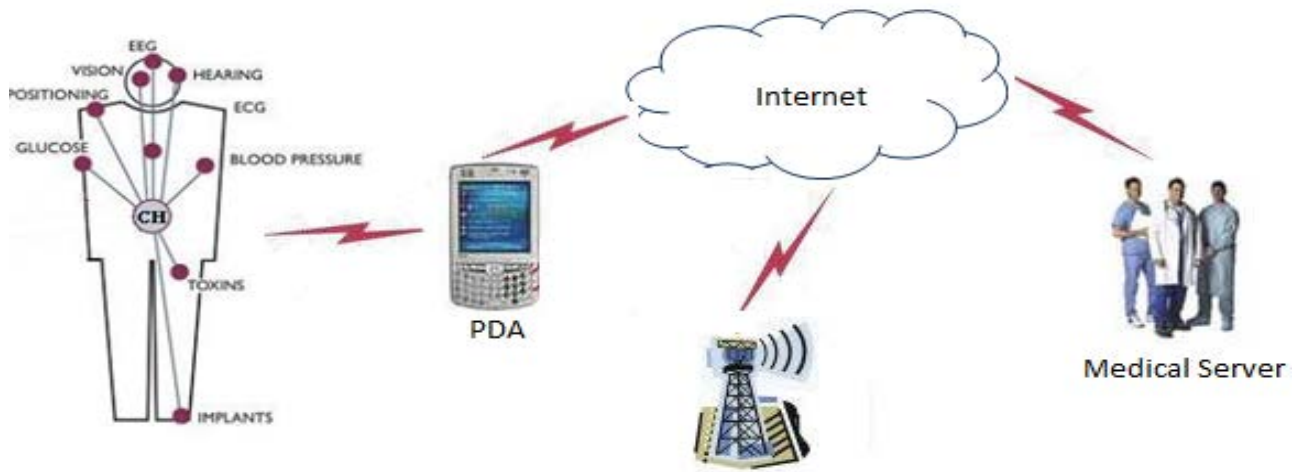

Fig. 4. General communication in BAN

## 4. We offer about BAN's communication

We show in figure 4 general communication in BAN and rest of the paper we explain about detail communication.

### 4.1. First case

One of the duties of CH, is that the individual sensors with mobile phones are considered in connection. Their relation to the way that CH vital signs such as temperature, pulse, blood pressure, blood sugar ... Humans to continuously control the body's vital signs, and if normal body's vital signs have been interrupted and is not happened in critical condition a long interval (eg once a day) by short-wavelength waves to phone and phone the person sends this information to the person via the Internet or through any other database sends personal information is stored in one file. If medical advice regarding this situation it was in the database as a phone message and sends the person himself can decide to act or not recommend it (freedom of the person he has not been forced to do work there no act.) insist on watching the arrival of these cases no information at the time specified does not exist if it did not have the phone does not delete the queue and send the information in specified intervals ($time_{01} < time_0$) sends. Figure 5 shows this case.

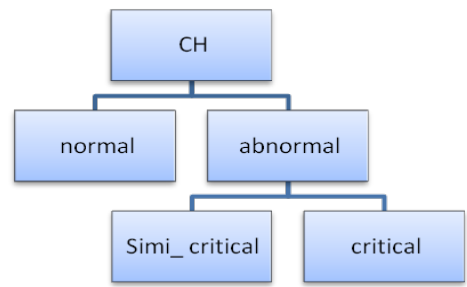

When entering the human body sensor in the database CH into a series of information that are unique for each person (for example fingerprints.) Main program when the body normally has repeatedly runs. (State of emergency has not occurred) and the program routinely done.

Data structure in the first case is as follows:

| fingerprints | CRC |
|---|---|

First abnormal format

| Fingerprint | sensor status | amounts of undesirable events | Tail |
|---|---|---|---|

First normal format

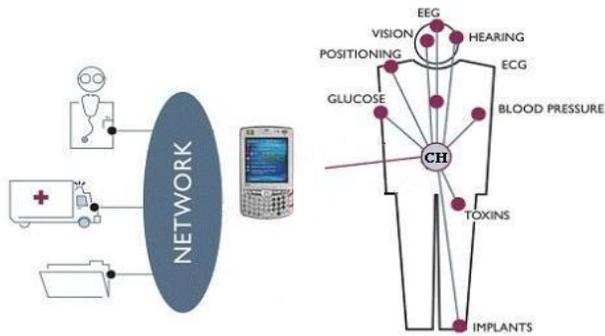

Fig. 5 . First case

Program runs in a normal situation:

While (the state of emergency has never occurred)
(Process (data from sensors)
   if (time system ==$time_0$)
  (
     Send (format data1)
     Send queue (format data1)
   )
  If (don't send & time system ==$time_{01}$)
  (
    If (inbox ~ = ACK)
    (
      Send queue
    )
   Else
   (
     Remove (queue)  )))

We can run program faster in normal times $time_0$ and $time_{01}$ If the interrupt hardware to draw until we spend more time processing CH is processing information received from sensors.

### 3.2. Second case

Now we consider the situation when a person is difficult, there is acute situation (in this case also interrupting the program runs continuously in the second stage of the program ends) and cell phone person is not available for this Send CH acute situation to the health care centers or emergency to Using GPRS phone and the person's status by individual service person to do.

 In this case, because we are in the residential area, if the CH , cell phone does not find the person. In this case, data to be released to all broadcasting for mobile phones that are around this person (eg a radius of 40 m) received (in this section assumes that they receive) and data to send that information by health care centers or emergency situations can be patient position using other people around the individual patient (eg, radius 40 m) to find Patient records by physicians, the individual decisions are correct. The figure 6 explains this case well. (Using the Document in the database).The latter data structure is as follows: Because most likely person in the event of such a case, and because residential energy consumption in this mode is lower for the sensor to use less energy to data instruction the above Such as Bluetooth to nearby devices will send. (Energy consumption for the sensor to the mobile phone is more important).

| fingerprint | CRC |
|---|---|
| | |

Second abnormal format

| bits emergency | Fingerprint | amounts of undesirable events | Tail |
|---|---|---|---|

Second normal format

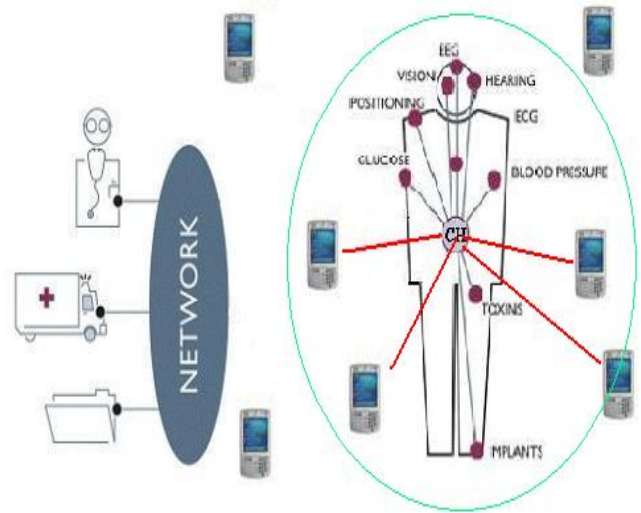

Fig .6. Second case

### 4.2.1. Bit of emergency

In this case, sensors that measure to submit such data to be nearby when the cell phones receive it without processing;

emergency bits removed and data downloaded to send to emergency centers and simultaneously send CH ACK. If CH does not receive answer the third case occurs. In this case, program execution stops at the second stage and then steps other cell phone will be sent. In fact, we have same interruption all cases.

### 4.3. Third case

Now consider the situation that we are not residential. (Eg, forest or mountains and desert interruption occurs ...) The individual is difficult. Acute situation occur and vital signs in the person gradually goes away and the person is not available for cell phones to CH this critical situation. This crisis will send to the emergency. We show this case in figure 7.In this case, because CH cell phone the person will not be the first CH data to be released to broadcast to all phones that are around this person (eg a radius of 40 m) and found to give and emergency center to inform (second case) CH because no answer from any other cell phone based on information received does not send its cluster head and decided to send the data with the same wavelength and energy production to phone and data Send to a database and database using position communications satellites or cellular towers to the nearest service center (emergency) to refer to individual service done[12,13]. (Because every person has a fingerprint data arrival is clear that the data belongs to whom)

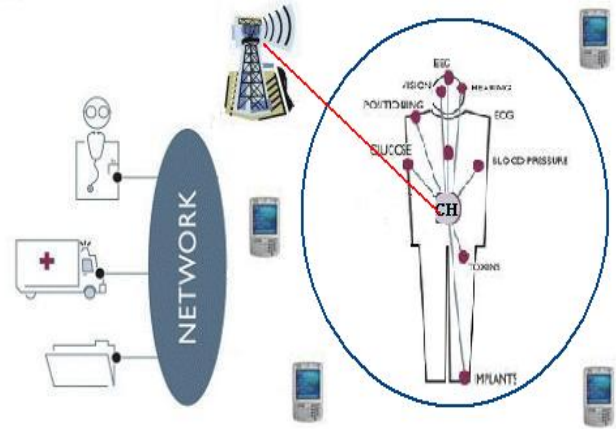

Fig. 7. Third case

Data structure in the third case as follows:

| Fingerprint | CRC |
|---|---|

Third abnormal format

| fingerprint | amounts of undesirable events | Tail |
|---|---|---|

Third normal format

### 5. Conclusions

Because each sensor depends on the performance of the neural network and neural network performance in such a way that its performance over time, which sees increased education and training neural networks in this paper continues and over time this leads to increased reliability neural network with time is the result of increased reliability is BAN.

 In this article, as far as we are energy sensors using wavelength range, we did a short addition to informing the border have given one hundred percent Vine increases reliability is as stable relationship with Given that communication via phone and through its pessimistic mode, the sensor is done. In addition, all decisions about that person's recommendations.

**Saeid Bahanfar** received the B.Sc. degree in Computer Software Engineering from Payam Noor University (PNU), Tabriz branch, Iran in 2008._ Currently, he is a M.Sc. student of Computer System Architecture in Islamic Azad University, Tabriz branch, Iran. His research interests include Residue Number System and VLSI Design, wireless sensor network, Neural network.

**Ladan Darouagarn** was born in Tabriz, Iran, on May 29, 1983. She received the B.Sc. degrees from University of Shabestar (Shabestar, Iran) and M.S.E. student in Islamic Azad University, Tabriz Branch in 2011. Her research interests are in the data aggregation in wireless sensor network. She is a member of Young Researchers Club.

**Helya Kousha** received her B.Sc. in Computer Software Engineering from Islamic Azad University, Shabestar branch, Iran in 2008. Currently, she is a M.Sc. student of Computer System Architecture in Islamic Azad University, Tabriz branch, Iran. Her main research interests include Computer Arithmetic, Residue Number System, wireless sensor network.

**Shahram Babaei** He is currently DR. student in Department of Computer Engineering of Science and Research Branch of Islamic Azad University, Tehran, Iran. His research interests are Computer Arithmetic ,Reliability in wireless sensor network, Cryptography, Network Security.